# Revisiting heat capacity of bulk polycrystalline $YBa_2Cu_3O_{7-\delta}$

Rajveer Jha, Poonam Rani and V.P.S Awana*
Quantum Phenomena and Applications Division, National Physical Laboratory (CSIR), New Delhi-110012, India

In this letter, we present the superconducting property characterization of a phase pure reasonably good quality $YBa_2Cu_3O_{7-\delta}$ sample. Studied compound is crystallized in orthorhombic Pmmm space group with lattice parameters *a*, *b*, and *c* are 3.829(2) Å, 3.887(1) Å and 11.666(3) Å respectively. Bulk superconductivity is observed below 90K as evidenced by resistivity and *dc*/*ac* magnetization measurements. The resistivity under magnetic field ($\rho TH$) measurements showed clearly both the intra-grain and inter-grain transitions, which are supplemented by detailed (varying frequency and amplitude) *ac* susceptibility studies as well. The upper critical field at 0K i.e., $H_{c2}(0)$ being determined from $\rho TH$ measurements with 50% criteria of resistivity drope is ~ 70 Tesla. Studied polycrystalline $YBa_2Cu_3O_{7-\delta}$ is subjected to detailed heat capacity ($C_P$) studies. $C_p$ exhibited well defined anomaly at below 90 K, which decreases with applied field. Though the $C_p$ anomaly/peak at $T_c$ reduces with applied field, the same is not completely suppressed in high applied fields of up to 12 Tesla. The Sommerfeld constant ($\gamma$) and Debye temperature ($\Theta_D$) as determined from low temperature fitting of $C_P(T)$ data to Sommerfeld-Debye model, are 10.65 mJ/mole-K$^2$ and 312.3 K respectively. The results are compared with existing literature on bulk polycrystalline superconducting $YBa_2Cu_3O_{7-\delta}$ sample

Key Words: YBa2Cu3O7 superconductor, *dc*/*ac* magnetic susceptibility, Magneto-transport and Heat capacity.

PACS Nos. 74.20.De, 74.25.Bt, 74.72.-h;

Dr. V. P. S. Awana, Principal Scientist
E-mail: awana@mail.npindia.org
Ph. +91-11-45609357, Fax-+91-11-45609310
Homepage www.fteewebs.com/vpsawana/

**Introduction**

Since the discovery of high-temperature superconductivity (*HTSC*) by Bednorz-Muller [1] at above 30K in rare earth metal oxide (LaBaCuO), various similar compounds were discovered with higher superconducting transition temperature ($T_c$). In this regards Superconducting $YBa_2Cu_3O_{7-\delta}$ (YBCO) system [2] with transition temperature above 90K is one of the most studied *HTSC* compound. Though several thousand articles had yet appeared on various *HTSC* compounds, yet their heat capacity analysis still lacks proper attention [3, 4]. This is primarily due to very higher upper critical fields of *HTSC* compounds. To distinguish between the electronic and phonon parts of heat capacity one needs to measure the superconductor in normal state below its superconducting transition temperature ($T_c$). This can be achieved only after applying a magnetic field higher than upper critical field of the superconductor. Because the upper critical field of *HTSC* cuprates is very high (> 100 Tesla), it is difficult to attain their normal state below $T_c$ and measure the low temperature pure electronic $C_p$. Another important aspect of *HTSC* cuprates is their granular nature, i.e. sandwiching of superconducting grains through the insulating grain boundaries. Polycrystalline superconducting samples can be considered as a system in which superconducting grains are weakly coupled with each other. Complex ac susceptibility ($\chi$) is useful in characterizing the granular nature of *HTSC* cuprates along with their resistivity and *dc* susceptibility. Magnetic irreversibility has already been reported in both *dc* magnetization and in *ac* susceptibility of *HTSC* [3, 5]. In these compounds $\chi$ possess both intrinsic (intra-grain) and coupling moments (inter-grain). The coupling component is very sensitive to temperature and applied amplitude of *ac* field. The loss component of the ac susceptibility has been used widely to probe the nature of weak links in polycrystalline samples [6]. The real and imaginary part of complex ac susceptibility represents two complementary aspects of flux dynamics in polycrystalline superconducting samples. It is well known that the real part of susceptibility consists of two transitions which correspond to the flux removal from intra-grain and inter-grain regimes. In accord, imaginary part contains two peaks which represent energy dissipation and ac losses due to the flux motion in intra-grain and inter-grain regions [6-12]. Study of resistivity under applied magnetic field ($\rho TH$) along with magnetization can reveal the microstructure of the superconductors. It is a very important tool for exploring the percolation nature between grains and grain boundary resistance. Inter and Intra grain connectivity and its impact on superconductivity can be understood. Basically the resistive transition occurs in two steps and interprets as the steep transition near the onset part, which is associated with superconductivity in individual grains, and the long transition tail is basically due to couplings regimes between grains or connective nature of grains [6-9]. Plotting the temperature derivative of resistivity data gives more insight to two steps structure of resistivity transition. It is essential to look at the temperature derivative of the resistivity in order to give a proper description of the superconducting transition [6]. Correspondingly temperature derivative of resistivity data gives narrow intense maxima approximately centered at $T_c$ and a broad peak at low temperatures [6-9]. The specific heat measurements along with resistivity and magnetization are been used, to determine the density of states (*DOS*) at Fermi level [13-15]. Low temperature specific heat fitting has been done earlier and two predictions were made for electronic specific heat of d-wave superconductivity, i.e., a $T^2$ term in zero field and an increased linear term in a magnetic field applied perpendicular to the $CuO_2$ planes [16, 17]. In this letter we present a detailed but crisp study of phase formation, resistivity under magnetic field ($\rho TH$), magnetization with temperature (*M-T*) and field (*M-H*) and specific heat $C_P(T)$ with and without magnetic field on a phase pure bulk polycrystalline $YBa_2Cu_3O_{7-\delta}$ sample.

**Experimental**

The $YBa_2Cu_3O_{7-\delta}$ sample is synthesized in air by solid-state reaction route. The stoichiometric mixture of $BaCO_3$, $Y_2O_3$, and CuO are ground thoroughly, calcined at 850ºC for 12h and then pre-sintered at 900ºC and 925ºC for 20h with intermediate grindings. Final sintered powder is palletized and sintered at 925ºC for 20h in air. Finally pellets are annealed in flowing oxygen at 650ºC for 12h and subsequently at 450ºC for 12h. The phase formation is checked with powder diffractometer, Rigaku (Cu-Kα radiation) at room temperature. The phase purity analysis and lattice parameter refining are performed by Rietveld refinement programme (Fullprof version). The resistivity, magnetization and Heat capacity measurements are carried out applying a field magnitude up to 12Tesla using Physical Properties Measurement system Quantum Designed *PPMS*-14Tesla.

**Results and discussion**

The studied $YBa_2Cu_3O_{7-\delta}$ sample is crystallized in single phase orthorhombic *Pmmm* space group without any impurities with in XRD limits, see Fig.1. This is confirmed from the Rietveld analysis of powder X-ray diffraction pattern shown in Fig. 1. The lattice parameters *a*, *b*, and *c* are 3.829(2) Å, 3.887(1) Å and 11.666(3) Å respectively. These lattice parameters are close to oxygen stoichiometric ($\delta = 0$) YBCO compound. The temperature dependent *dc* magnetic susceptibility is shown in the Fig 2. The measurement is carried out at 10Oe applied magnetic field down to 5 K in both Zero Field Cooled (*ZFC*) and Field Cooled (*FC*) situations. The diamagnetic signal starts from 91 K in both *ZFC* and *FC* magnetization, indicating the establishment of superconductivity below this temperature. Further, a clear indication for flux pinning of the sample is evident from the separation of *FC* and *ZFC* signals or in other words reduction of Meissner fraction (ratio of field cooled to zero fielded cooled magnetization). The studied YBCO sample show reasonable shielding (60%) and 20% Meissner volume fractions. Imaginary part of *ac* Susceptibility of the sample measured at different *ac* field amplitudes (1-17 Oe) with zero bias *dc* fields is given in inset-I of Fig. 2. The imaginary part contains two peaks primarily due to the flux motion in (a) intra-grain and (b) inter-grain regions, representing the energy dissipation and *ac* losses in the superconducting sample. The high temperature peak amplitude in the imaginary part is associated with individual superconducting grains and is a measure of grain size [9, 10]. The high temperature peak (intra-grain) is almost insensitive to the applied field, while the low temperature peak (inter-grain couplings) is highly sensitive to the field amplitudes. The inter-grain peak lies very close to the intra-grain peak at lower amplitudes but shifts to the low temperatures at higher amplitudes. It can be say that at the higher amplitudes the field penetrates the sample deeper than in the lower fields. Due to persisting strong inter-grain couplings it is difficult to distinguish intra-grain and inter granular regions in lower amplitudes. Worth mentioning is the fact, that the result shown in inset I of Fig. 2 is distinct in comparison to the earlier reports [6-12]. The very clear appearance of both intra and inter grain peaks in current sample is remarkable. In fact to achieve such nice data one needs to make sure to get absolute zero *dc* field situation before embarking on *ac* susceptibility measurements. Inset-II shows the isothermal magnetization (*M-H*) at various temperatures below $T_c$ for the studied YBCO sample. The isothermal magnetization loops (*M-H*) loops are taken up to 80 kOe at 5, 20, 50 and 70 K temperatures of the $YBa_2Cu_3O_{7-\delta}$ sample. The lower critical field ($H_{c1}$) value is around 2300 Oe at 5 K. The critical current density is calculated from the *M-H* loops with the help of Bean's model [18], which is typically of the order of $10^5$ A/cm$^2$ at 0 field and 5 K. Figure 3 shows the temperature dependent resistivity under magnetic field (up to 13 Tesla), of the studied $YBa_2Cu_3O_{7-\delta}$ sample. In zero fields, the $T_c$ of sample is around 90 K. The $\rho(T)H$ curve shows basically two transitions, i.e. the onset of resistivity drop and the exact $\rho = 0$ state. It can be seen that the effect of magnetic field is weaker at the onset part near the normal state in comparison to the tail part close to $\rho = 0$ state. Also, the offset of $T_c$ ($\rho = 0$) is moved to lower temperatures with increasing field. This occurs near the onset part, where superconductivity persists only inside individual grains and the superconducting fraction is quite small. A long range superconducting state with zero resistance is achieved by means of a percolation like process that overcomes the weak links between grains. The broadening of the tail part occurs due to the anisotropic nature and the disturbances in the percolation path between grains. It is well known that long-range superconducting state with zero resistance is achieved by means of a percolation like process that overcomes the weak links between grains [19]. Temperature dependency of resistive upper critical field, $H_{c2}$ ($T$), using midpoint data criteria, where the resistivity is half of its normal state value, is shown in inset of Fig.3. The $H_{c2}$ ($0$) values is above 70 Tesla, which is in agreement with earlier studies [20 - 22].

The temperature dependence of specific heat $Cp(T)$ being measured from 2-250 K in zero field is shown in Fig.4. At around superconducting temperature ($T_c = 90$ K) the jump in specific heat is clearly visible in zero magnetic field. The value of the jump is found to be 139.7 J mol$^{-1}$ K$^{-1}$ which can be seen clearly from the inset of the Fig.4. Inset of the Fig.4 shows the specific heat in various applied fields 0, 1, 3, 5 and 12 Tesla. It is noticed that with increasing applied field both the $T_c$ and the transition jump height decreases monotonically. Figure 5 shows the Plots of $C_p/T$ versus $T^2$ for applied field 0, 1, 3, 5 and 12 Tesla. The inset of Fig. 5 shows the zoom part of $C_p/T$ versus $T^2$ near $T_c$. The $C_p(T)H$ anomaly/peak near $T_c$ is seen more pronouncedly in $C_p/T$ versus $T^2$ than the $Cp(T)$ plots shown in Fig. 4. It is clear from inset of Fig. 5 that the $Cp$ anomaly is not speared completely smeared even in highest applied field of 12 Tesla. Figure 6 shows the low temperature normal-state $C_p(T)$ data from 2-20 K for $H = 12$ Tesla which has been fitted to the Sommerfeld–Debye expression

$$C(T) = \gamma T + \beta T^3 + \delta T^5 \ldots\ldots\ldots\ldots\ldots\ldots\ldots\ldots\ldots\ldots \quad (1)$$

where the $\delta T^5$ term represents the anharmonic contribution. From this fitting the values of Sommerfeld constant ($\gamma$) and $\beta$ are obtained. The $\gamma$ and $\beta$ give the value of electronic density of states and approximate value of Debye temperature respectively. The values obtained are $\gamma$ = 10.65 mJ mol$^{-1}$ K$^{-2}$, $\beta$ = 6.73 mJ mol$^{-1}$ K$^{-4}$ and $\delta$ = 0.0003 mJ mol$^{-1}$ K$^{-6}$. The fitting being shown in figure 6 contains the $T^3$ contribution to $C_p$, which originates from phonons. Inset of the Fig.6 shows the electronic specific heat anomaly in zero magnetic fields at around 90 K. The electronic specific heat ($C_{es}$) is calculated by subtracting the normal state fitted $C_p(T)$ from the experimental raw data. The normalized value of jump ($C_{es}/ \gamma T$) is found to be above 3.0, which is more than twice in comparison to the Bardeen–Cooper–Schrieffer (*BCS*) value of 1.43. The linear term $\gamma T$ is found in low temperature $C_p$ of normal metals due to the contributions from the near the Fermi surface electrons. [17, 18]. It is difficult to see this contribution at higher temperatures because at these temperatures the phonon contribution dominates. The observation of a linear term in the specific heat at low temperatures, if intrinsic, would represent a departure from the BCS prediction and hence is of considerable interest. The finite value of $\gamma$ indicates finite electronic density of states at low energy, in zero applied field. It has also been reported that the large value for $\gamma$ may originate from a disorder-generated finite density of quasiparticle states near the d-wave nodes [23]. The Debye temperature is calculated by using $\Theta_D = (234zR/\beta)^{1/3}$, here $z$ being number of atoms per formula unit and $R$ is the gas constant. Taking the fitted value of $\beta$ from eqn. 1, the calculated value of $\Theta_D$ is 312.3 K, which is close to the reported values [15]. The fitted value of Sommerfeld constant ($\gamma$ = 10.65 mJ/K$^2$mol) is used to calculate the value of electronic density of states at the Fermi level $N$ ($E_F$) using the formula $N(E_F) = 3\gamma/\pi^2 K_B^2$ The value of $N(E\mathrm{F})$ is calculated as 3.312 states/eV f.u. Henceforth it is concluded that we observed well defined $C_p(T)$ peak in a bulk polycrystalline YBCO superconductor near its $T_c$ of 90 K. Though the $C_p(T)$ peak could not be completely smeared off with highest applied magnetic field of 14 Tesla, yet the resultant low temperature electronic heat capacity is fitted with known equations and reasonable electronic parameters are obtained. In conclusion, a brief note is presented on superconducting properties of reasonably good quality bulk polycrystalline $YBa_2Cu_3O_{7-\delta}$ superconductor. The structural, electrical, magnetic and thermal characterization is summarized briefly in crisp manner.

**References**

1. J.G. Bednorz, K.A. Muller, Zeitschrift für Physik B Condensed Matter **64** 189 (1986).
2. M.K. Wu, J.R. Ashburn, C.J. Tong, P.H. Hor, R.L. Meng, L. Gao, Z.H. Huang, Y.O. Wang, C.W. Chu, Physical Review Letters **58** 908 (1987).
3. M. Roulin, A. Junod, E. Walker, Physica C, 260, 257 (1996).
4. V.G. Bessergenev, Yu.A. Kovalevskaya, V.N. Naumov , G.I. Frolova, PhysicaC, 245, 36 (1995).
5. D. X. Chen, E. Pardo, A. Sanchez, and E. Bartolome, APPLIED PHYSICS LETTERS **89**, 072501 (2006).
6. A. A. Bahgat, E. E. Shaisha, M. M. Saber, Physica B 399 70 (2007).
7. Rajvir Singh, R. Lal, U. C. Upreti, D. K. Suri, A.V. Narliker and V. P. S. Awana, Phys. Rev. B 55, 1216 (1997).
8. S. L. Shinde, J. Morrill, D. Goland, D. A Chance and T. McGuire, Phys. Rev. B 41, 8838 (1989).
9. Roland V. Sarmago and Bess G. Singidas, Supercond. Sci. Technol. 17, S578 (2004).
10. D. X. Chen, R. B. Goldfarb, J. Nogues and K. V. Rao, J. Appl. Phys. 63(3), 98 (1988).
11. A. K. Grover, C. Radhakrishnamurthy, P. Chaddah, G. Ravikumar, and G. V. Subba Rao, Pramana J. Phys. 30, 569 (1988).
12. I. E. Edmond and L. D. Firth, J. Phys. Condens. Matter 4, 3813(1992).
13. V.N. Naumov, Phys. Rev. B 49 13247(1994).
14. J.W. Loram, K. A. Mirza, J. R. Cooper, and W. Y. Liang, Phys. Rev. Lett. **71**, 1740 (1993).

15. Yuxing Wang, Bernard Revaz, Andreas Erb, and Alain Junod, Phys. Rev. B **63**, 094508 (2001).

16. N. E. Phillips, Phys. Rev. **114**, 676 (1959).

17. Kathryn A. Moler, David L. Sisson, Jeffrey S. Urbach, Malcolm R. Beasley, Aharon Kapitulnik, David J. Baar, Ruixing Liang, and Walter N. Hardy Phys. Rev. B, 55, 3954 (1997).

18. N.P. Liyanawaduge, Anuj Kumar, Rajveer Jha, B.S.B. Karunarathne, V.P.S. Awana, Journal of Alloys and Compounds 543 135, (2012).

19. S. Dadras, Y. Liu, Y.S. Chai, V. Daamehr, K.H. Kim, Physica C: Superconductivity 469 55 (2009).

20. P.J.M. Van-Bentum, H.V. Kempen, L.E.C. Van-de-Leemput, J.A. Perenboom, L.W.M. Schreurs, P.A.A. Teunissen, Physical Review B 36 5279 (1987).

21. U.D.S. Kalavathi, T.S. Radhakrishnan, G.V.S. Rao, Journal De Physique C 8 2167 (1988).

22. O. Labrode, J.L. Tholence, P. Lejay, A. Sulpice, R. Tournier, J.L. Capponi, C. Michel, J. Provost, Solid State Communications 63 877 (1987).

23. Scott C. Riggs, O. Vafek, J. B. Kemper, J. B. Betts, A. Migliori, F. F. Balakirev, W. N. Hardy, Ruixing Liang, D. A. Bonn and G. S. Boebinger, Nature Physics 7, 332, (2011).

**Figure Captions:**

**Figure 1:** Observed (*open circles*) and calculated (*solid lines*) *XRD* patterns of $YBa_2Cu_3O_{7-\delta}$ sample at room temperature.

**Figure 2:** *DC* magnetization (both *ZFC* and *FC)* plots for $YBa_2Cu_3O_{7-\delta}$ sample measure in the applied magnetic field, H= 10 Oe. Inset-I show the imaginary part of ac susceptibility and Inset-II shows the MH curve for the same sample at different temperature 5-70 K.

**Figure 3:** Temperature dependence of the resistivity $\rho(T)$ under magnetic fields 0 - 13 T for the $YBa_2Cu_3O_{7-\delta}$. Inset shows the upper critical field $H_{c2}$ found from 50% of the resistivity $\rho_n(T)$ for the same samples.

**Figure 4:** Specific heat versus temperature plot $C_p$ $(T)$ in the temperature range of 2–250 K for the studied $YBa_2Cu_3O_{7-\delta}$ sample, insets show the $C_p$ *vs* $T$ in the temperature range of 84–94 K under applied fields 0, 1, 3, 5, and 12 Tesla.

**Figure 5:** $C_p/T$ versus $T^2$ under the applied fields 0, 1, 3, 5, and 12 Tesla for the $YBa_2Cu_3O_{7-\delta}$, inset is zoom part near jump in specific heat capcity.

**Figure 6:** Fitted curve of *Cp* versus $T$ for the applied field 12 Tesla for the $YBa_2Cu_3O_{7-\delta}$, the insets show the electronic specific heat anomaly at $T_c$.

**Fig. 1**

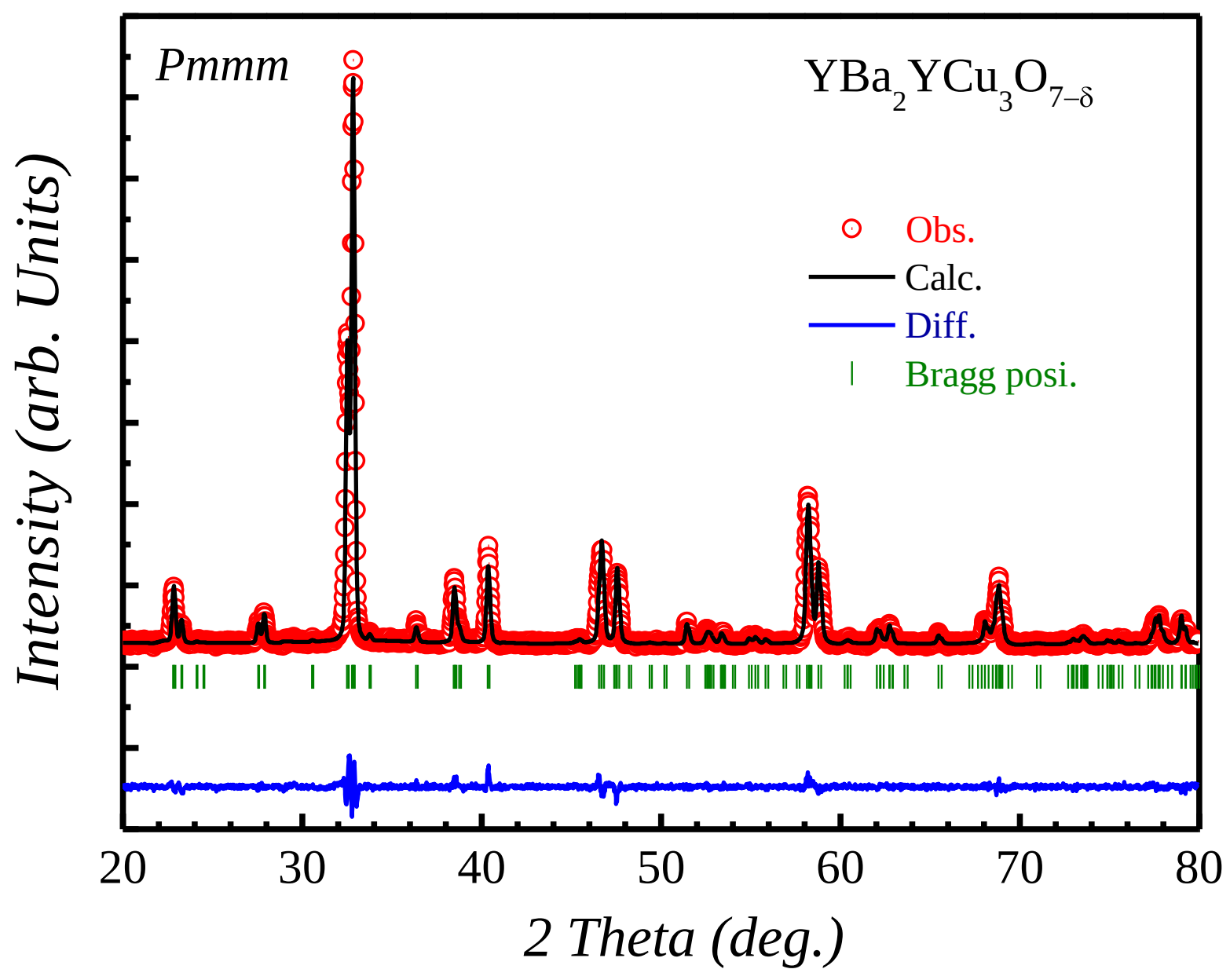

**Fig. 2**

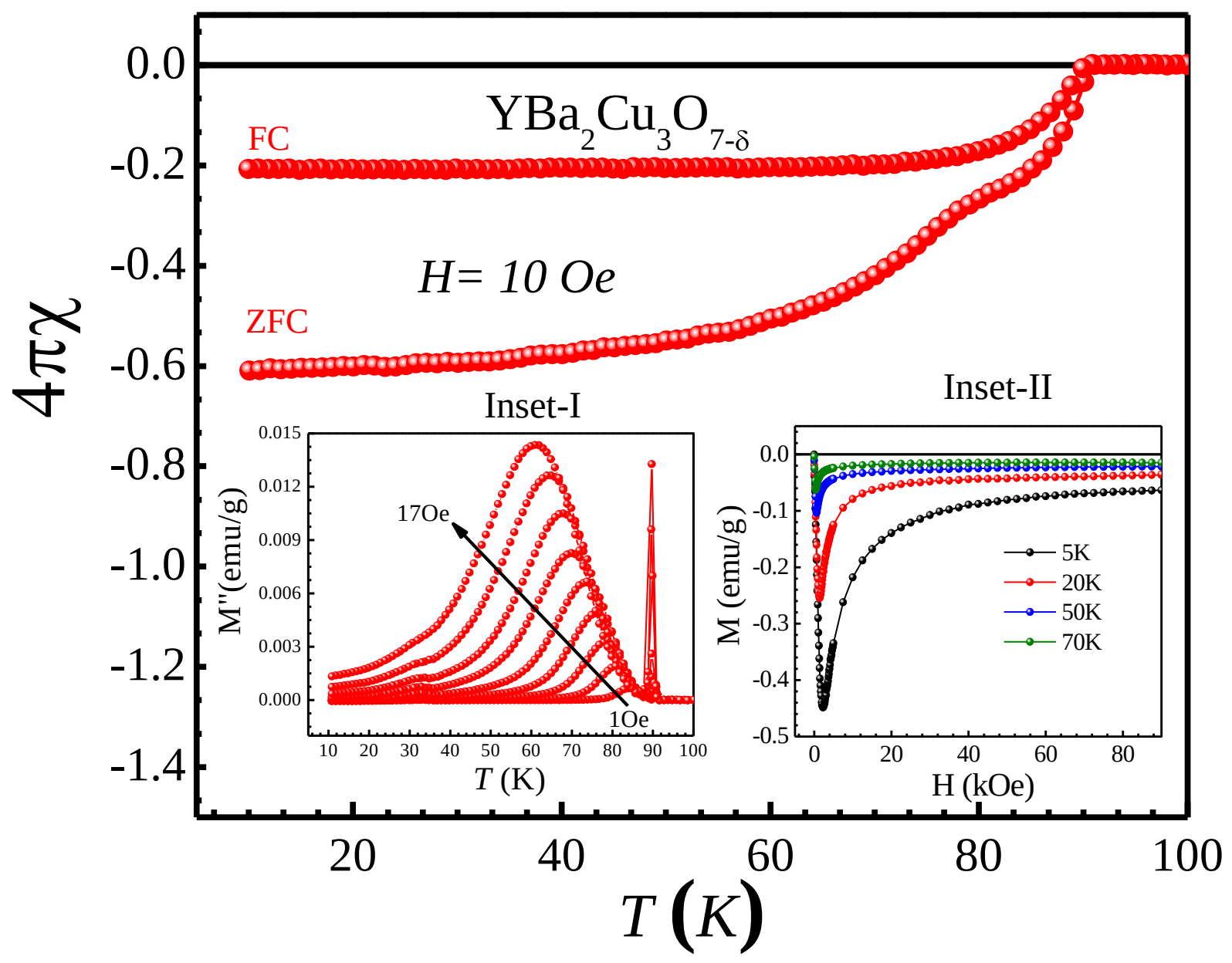

**Fig. 3**

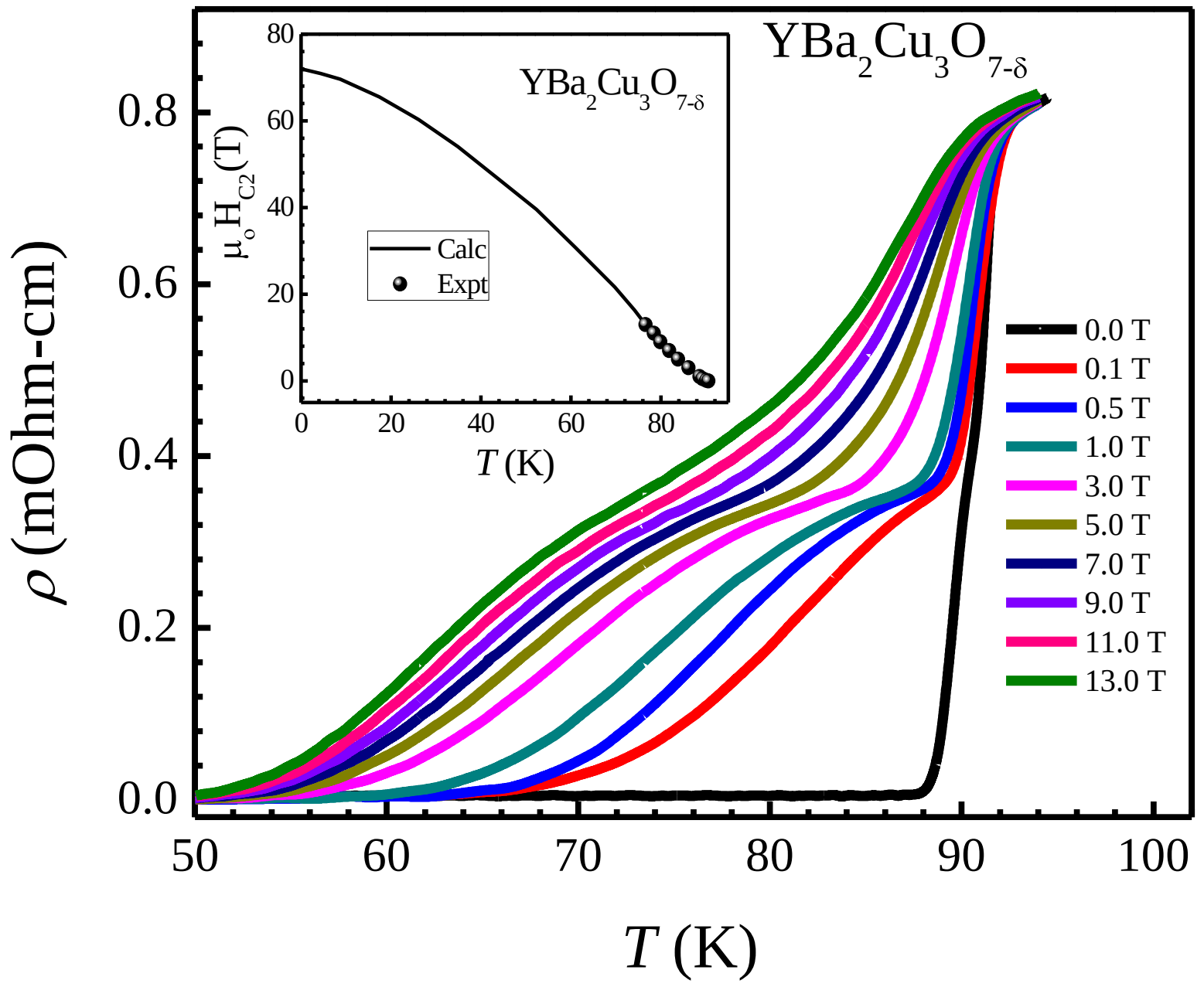

**Fig. 4**

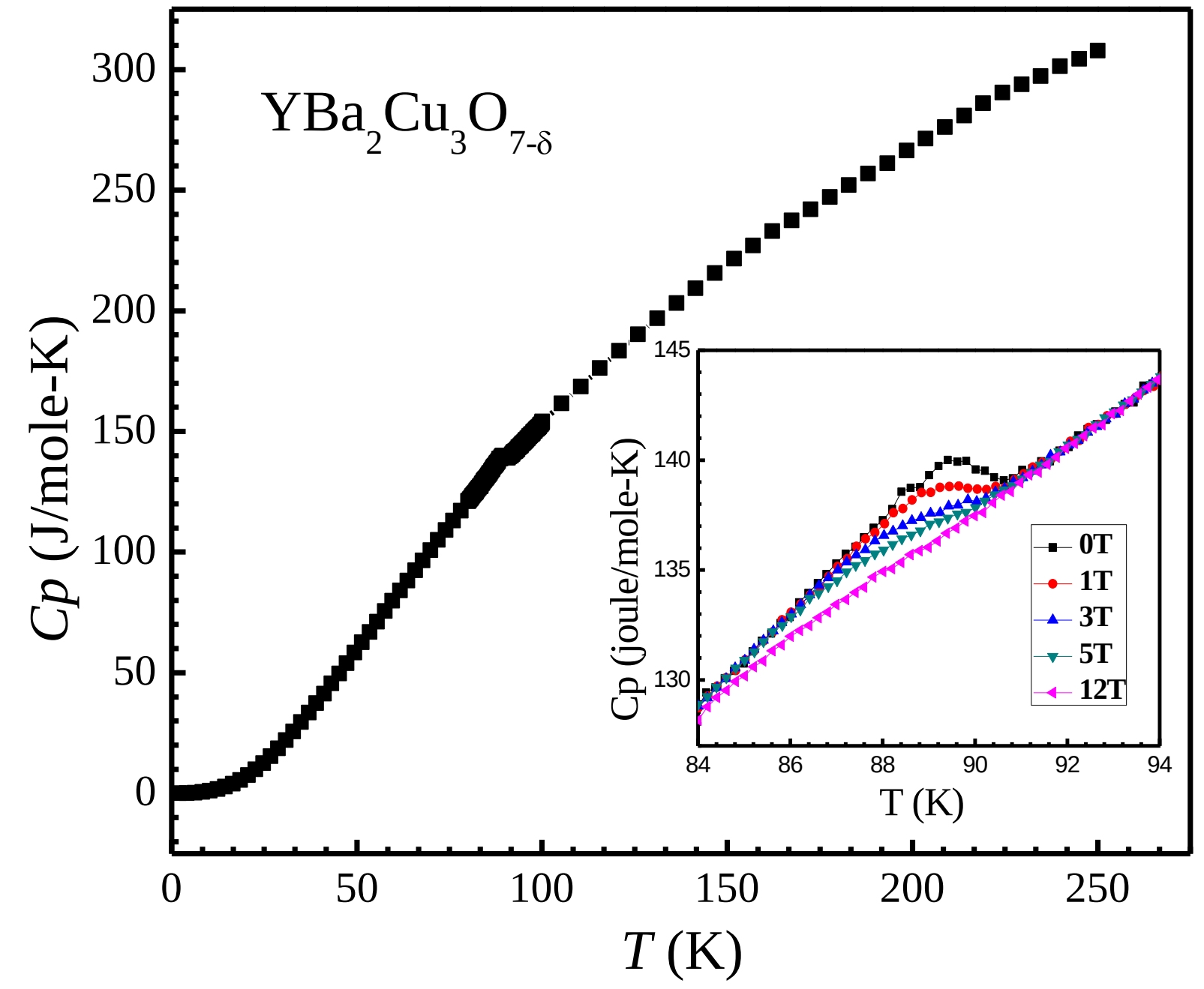

**Fig. 5**

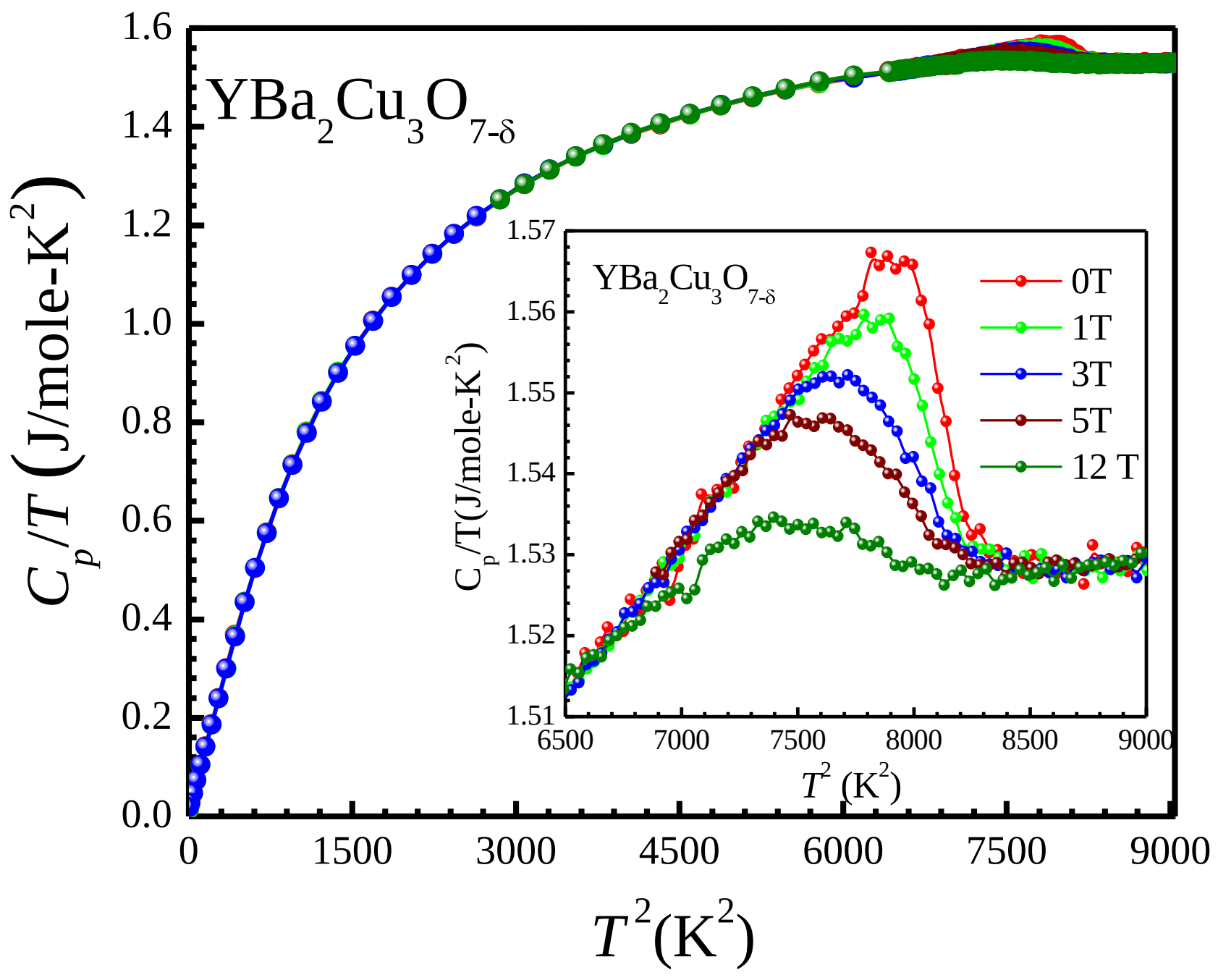

**Fig. 6**

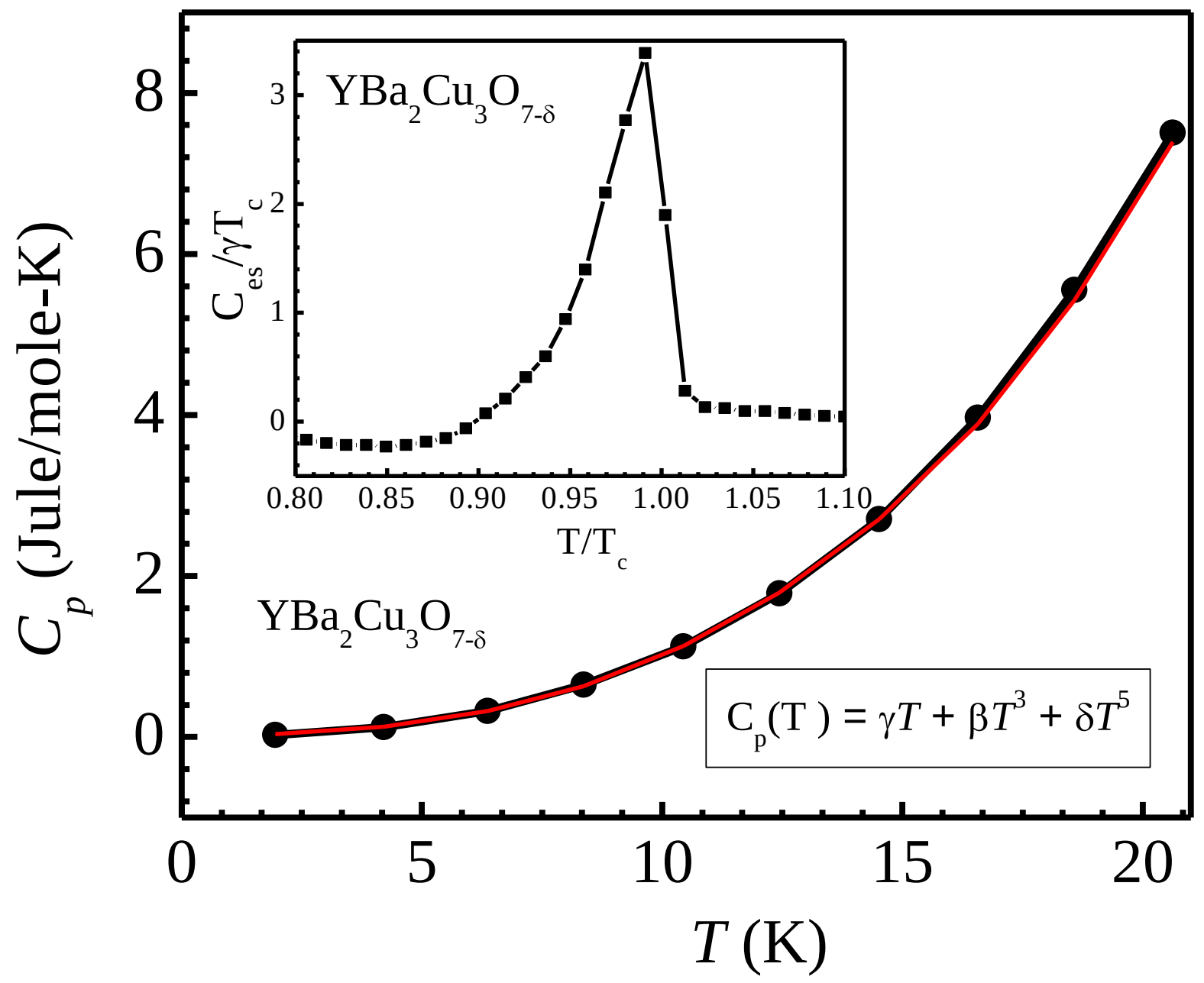

YBa2Cu3O7-δ
Ces/γTc
T/Tc
0.80
0.85
0.90
0.95
1.00
1.05
1.10
0
1
2
3
YBa2Cu3O7-δ
Cp(T ) = γT + βT3 + δT5
Cp (Jule/mole-K)
T (K)
0
5
10
15
20
0
2
4
6
8